\pdfoutput=1
\documentclass[sigconf]{acmart}

\usepackage{booktabs} 
\usepackage{xspace}
\usepackage{caption}
\usepackage{enumitem}
\usepackage{subcaption}

\definecolor[named]{Green}{cmyk}{0.8,0,1,0.19}

\usepackage{pgfplotstable}
\usepackage{pgfplots}
\definecolor{darkgreen}{rgb}{0,0.5,0.17}

\newcommand*{\figstepstikzscale}{0.8} 
\newcommand*{\plotscale}{0.70} 
\newcommand*{\belowplotvspace}{-10pt} 

\usepackage[colorinlistoftodos,prependcaption,textsize=tiny]{todonotes}

\graphicspath{ {figures/} } 

\newcommand{\tool}[1]{\textsc{#1}\xspace}

\newcommand{\clazz}[1]{\texttt{#1}\xspace}
\newcommand{\method}[1]{\texttt{#1}\xspace}
\newcommand{\prog}[1]{\mbox{\texttt{#1}}\xspace}
\newcommand{\openjdk}{\textsc{OpenJDK}\xspace}
\newcommand{\jdk}{\textsc{JDK}\xspace}
\newcommand{\java}{\textsc{Java}\xspace}
\newcommand{\afl}{\tool{AFL}}

\newcommand{\kelinci}{\tool{Kelinci}}
\newcommand{\kelinciwca}{\tool{KelinciWCA}}
\newcommand{\badger}{\tool{Badger}}

\newcommand{\commonscompress}{\tool{Apache Commons Compress}}
\newcommand{\asm}{\tool{ASM}}
\newcommand{\jpf}{\tool{JPF}}
\newcommand{\javapathfinder}{\tool{Java PathFinder}}
\newcommand{\spf}{\tool{SPF}}
\newcommand{\symbolicpathfinder}{\tool{Symbolic PathFinder}}
\newcommand{\symexe}{\emph{SymExe}\xspace} 
\newcommand{\darpa}{\textsc{DARPA}\xspace}

\usepackage{listings}

\usepackage{balance}

\setcopyright{none}
\acmYear{2018} 
\acmConference[ISSTA'18]{27th ACM SIGSOFT International Symposium  on  Software Testing and Analysis}{July 16--21, 2018}{Amsterdam, Netherlands}
\acmDOI{10.1145/3213846.3213868}
\acmISBN{}

\begin{document}
\title{Badger: Complexity Analysis with Fuzzing and Symbolic Execution}

\author{Yannic Noller}
\affiliation{%
  \institution{Humboldt University of Berlin}
  \city{Berlin}
  \country{Germany}
}
\email{noller@informatik.hu-berlin.de}

\author{Rody Kersten}
\affiliation{%
  \institution{Synopsys, Inc.}
  \city{San Francisco}
  \country{USA}
}
\email{rody@synopsys.com}

\author{Corina S. P\u{a}s\u{a}reanu}
\affiliation{%
  \institution{Carnegie Mellon University Silicon Valley, NASA Ames Research Center} 
  \city{Moffet Field}
 \country{USA}
}
\email{corina.s.pasareanu@nasa.gov}

\renewcommand{\shortauthors}{Yannic Noller, Rody Kersten, Corina S. P\u{a}s\u{a}reanu}

\begin{abstract}
  Hybrid testing approaches that involve fuzz testing and symbolic execution have shown promising results in achieving high code coverage, uncovering subtle errors and vulnerabilities in a variety of software applications.
In this paper we describe Badger - a new hybrid approach for complexity analysis, with the goal of discovering vulnerabilities which occur when the worst-case time or space complexity of an application is significantly higher than the average case. 

Badger uses fuzz testing to generate a diverse set of inputs that aim to increase not only coverage but also a resource-related {\em cost} associated with each path. Since fuzzing may fail to execute deep program paths due to its limited knowledge about the conditions that influence these paths, we complement the analysis with a symbolic execution, which is also customized to search for paths that increase the resource-related cost. Symbolic execution is particularly good at generating inputs that satisfy various program conditions but by itself suffers from path explosion. Therefore, Badger uses fuzzing and symbolic execution in tandem, to leverage their benefits and overcome their weaknesses.  

We implemented our approach for the analysis of Java programs, based on Kelinci and  Symbolic PathFinder. We evaluated Badger on Java applications, showing that our approach is significantly faster in generating worst-case executions compared to fuzzing or symbolic execution on their own.

\end{abstract}


\maketitle

\section{Introduction}
\label{sec:intro}

In recent years, fuzz testing has emerged as one of the most promising testing techniques for finding correctness bugs and security vulnerabilities in software. It is used routinely by major software companies such as Microsoft~\cite{MicrosoftSecurity,Godefroid:2012:SWF:2090147.2094081} and Google~\cite{Google}. While a large fraction of the inputs generated with fuzzing may be invalid, it can be more effective in practice than more sophisticated testing techniques --~such as the ones based on symbolic execution~\cite{Godefroid:2005:DDA:1065010.1065036,Sen2005}~--
due to the low computation overhead involved in fuzzing.

Fuzz testing tools, such as \afl~\cite{afl} and \tool{LibFuzzer}~\cite{libfuzzer}, have proven very successful by finding bugs and vulnerabilities in a variety of applications, ranging from image processors and web browsers to system libraries and various language interpreters. For example, \afl was instrumental in finding several of the Stagefright vulnerabilities in Android, the Shellshock related 
vulnerabilities in \tool{BIND} 
as well as numerous bugs in 
popular applications and libraries such as \tool{OpenSSL}, \tool{OpenSSH}, \tool{GnuTLS}, \tool{GnuPG}, \tool{PHP}, \tool{Apache}, and \tool{IJG jpeg}. 
In a nutshell, \afl uses genetic algorithms to {\em mutate} user-provided inputs using byte-level operations. These mutations are guided by coverage information obtained from running the analyzed program on the generated inputs. The interesting mutants (that are shown to increase coverage) are saved and mutated again. The process continues with the newly generated inputs, with the goal of generating a diverse set of inputs that increase the coverage of the program.

Motivated by the success of fuzz testing, we explore here the application of the technique to algorithmic complexity analysis. Characterizing the algorithmic complexity of a program has many practical applications as it enables developers to reason about their programs, understand performance bottlenecks and find opportunities for compiler optimizations. Algorithmic complexity analysis can also reveal worst-case complexity vulnerabilities, which occur when the worst-case time or space complexity of an application is significantly higher than the average case. In such situations, an attacker can mount Denial-of-Service attacks by providing inputs that trigger the worst-case behavior, thus preventing benign users to use the application.


There are several challenges in adapting fuzz testing to algorithmic complexity analysis.
First, fuzz testers like \afl are designed to generate inputs that increase code coverage, while for complexity analysis one is interested in generating inputs that trigger worst case execution behavior of the programs. Furthermore, fuzzers are known to be good at finding so called {\em shallow} bugs but they may fail to execute deep program paths \cite{DBLP:conf/ndss/StephensGSDWCSK16}, i.e. paths that are guarded by specific conditions in the code.  This is due to the fact that the fuzzers have little knowledge about which inputs affect which condition in the code. On the other hand, symbolic execution techniques are particularly well suited to find such cases, but usually are much more expensive in terms of computational resources required.

We therefore propose an analysis tool that uses fuzzing and symbolic execution in tandem, to enable them to find worst case program behaviors, while addressing their limitations. Specifically, we present \badger: a framework that combines fuzzing and symbolic execution for automatically finding algorithmic complexity vulnerabilities in Java applications. 

Our hybrid approach works as follows. We first run a fuzzer to generate a diverse set of inputs. 
For fuzzing, we build on \kelinci~\cite{ccs2017}, an \afl-based fuzzer for \java programs. We modify \kelinci and \afl to add a new heuristic
to account for resource-usage {\em costs} of program executions, meaning that the inputs generated by the fuzzer are marked as important if they obtain either an increased execution cost or new coverage. We call this tool \kelinciwca. The cost is defined in terms of 
number of conditions executed, actual execution time as well as user-defined costs that allow us to keep track of memory and disk usage as well as other resources of interest particular to an application.

The inputs generated by the fuzzer may cover a large set of executions but may fail to exercise deep program behavior. 
This can happen because of some hard-to-solve conditions that guard deep executions, as discussed above. 
At some user-defined point in time, the inputs are transferred to the symbolic execution side which analyzes them with the goal of 
producing new inputs that increase the cost and/or the coverage. These inputs are passed back to the fuzzer and 
the process continues until a vulnerability is found or a user-defined threshold is met.

For symbolic execution we use \symbolicpathfinder~(\spf), a symbolic execution tool for \java bytecode~\cite{Pasareanu:2013:ASE}. We modified \spf by adding a mixed concrete-symbolic execution mode, similar to concolic execution~\cite{Sen2005} which allows us to {\em import} the inputs generated on the fuzzing side and quickly reconstruct the symbolic paths along the executions triggered by the concrete inputs. These symbolic paths are then organized in a {\em tree} which is analyzed with the goal of generating new inputs that expand the tree. The analysis is guided by novel {\em heuristics} on the SPF side that favor new branches that increase resource-costs. 
The newly generated inputs are passed back to the fuzzing side.

A novelty of our approach is the handling of user-dependent costs, which get translated into symbolic costs on the symbolic execution side and are handled by running a symbolic maximization procedure, to generate the worst-case inputs. This broadens the application of \badger over previous symbolic execution techniques~\cite{wise,Luckow2017}, which could only handle  simple, concrete costs. 

Scalability is achieved in two ways. First, constraint solving is turned off during concolic execution, and is used only for generating new inputs, when expanding the tree. This is done selectively, guided by the heuristics. Furthermore, the tree is saved in memory, and expanded incrementally, only when new inputs are generated. 

We demonstrate how \badger is able to find a large number of inputs that trigger worst-case complexity in complex applications, and we show that it performs better than its parts (i.e. fuzzing and symbolic execution separately).

We note that we are not the first to use fuzzing and symbolic execution in a complementary manner. Tools such as \tool{Mayhem}~\cite{Cha:2012:UMB:2310656.2310692} and \tool{Driller}~\cite{DBLP:conf/ndss/StephensGSDWCSK16} are prominent examples. \tool{Mayhem} won first place at the recent DARPA Cyber Challenge~\cite{darpa_cgc_archive}, and \tool{Driller} later matched those results. There are many other similar hybrid approaches, which we discuss in Section~\ref{sec:relwork}. However, all these previous hybrid approaches aim to increase code coverage and we believe that we are the first to explore this combination for complexity analysis. The recent work on \tool{SlowFuzz}~\cite{Petsios2017} explores fuzzing for worst-case analysis (WCA). Although that work addresses binaries (and not \java) and uses a different fuzzer~\cite{libfuzzer}, it is similar in spirit to our \kelinciwca tool. For this reason, we include in our experiments \java versions of the same (or similar) examples as in \cite{Petsios2017}. While we can not compare \badger to \tool{SlowFuzz} directly, we compare \badger with \kelinciwca, which should give an indication whether a combination of \tool{SlowFuzz} with symbolic execution could achieve similar benefits.





\section{Background}

\label{sec:background}
\subsection{Fuzz Testing Java Programs with Kelinci}
\label{sec:kelinci}

\kelinci
is an interface to execute \afl on \java programs~\cite{ccs2017}.  It adds \afl-style instrumentation to Java programs and communicates results back to a simple C program that interfaces with the \afl fuzzer. 
This in turn behaves as a C program that was instrumented by one of \afl's compilers.

The first step when applying \kelinci is to add \afl-style instrumentation to a Java program. \afl uses a 64 kB region of shared memory for communication with the target application. Each basic block is instrumented with code that increments a location in the shared memory bitmap corresponding to the branch made into this basic block.
The Java version of this instrumentation is the following, which amounts to 12 bytecode instructions after compilation:

\begin{verbatim}
Mem.mem[id^Mem.prev_location]++;
Mem.prev_location = id >> 1;
\end{verbatim}

In this example, the \clazz{Mem} class is the Java representation of the shared memory and also holds the (shifted) \prog{id} of the last program location. The \prog{id} of a basic block is a compile-time random integer, where $0 \leq id < 65536$ (the size of the shared memory bitmap). The idea is that each jump from a block $id1$ to a block $id2$ is represented by a location in the bitmap $id1 \oplus id2$. While obviously there may be multiple jumps mapping to the same bitmap location, or even multiple basic blocks which have the same \prog{id}, such loss of precision is considered rare enough to be an acceptable trade-off for efficiency. The reason that the \prog{id} of the previous location is shifted is that, otherwise, it would be impossible to distinguish a jump $id1 \rightarrow id2$ from a jump $id2 \rightarrow id1$. Also, tight loops would all map to the location 0, as $id \oplus id = 0$ for any $id$.
Instrumentation is added to the program at compile time using the \asm bytecode manipulation framework~\cite{asm}. 

Based on this lightweight instrumentation, \afl will prioritize input files that lead to newly covered branches as ancestors for the next generation of input files. \kelinci was used to find bugs in various Java applications, including \tool{Apache Commons Imaging} and \tool{OpenJDK 9}~\cite{ccs2017}.

\subsection{Symbolic Execution and Symbolic PathFinder}
\label{sec:spf}

Symbolic execution~\cite{King:1976:SEP:360248.360252,Clarke:1976:SGT:1313320.1313532}
is a program analysis technique which executes programs on symbols in place of concrete inputs. When a decision is encountered, all branches are explored. Branch conditions are aggregated into a \emph{path condition}, a constraint over the symbolic program inputs. Solving the path condition using an off-the-shelf solver (e.g. \tool{Z3}~\cite{DeMoura:2008:ZES:1792734.1792766}) can detect infeasible paths as well as generate actual inputs (solutions of the constraint) that lead execution down the corresponding path. A typical use-case for symbolic execution is test-case generation~\cite{Clarke:1976:SGT:1313320.1313532,Cadar2008,Braione:ISSTA:2017}. In many cases, it can also detect faults directly~\cite{Godefroid:2012:SWF:2090147.2094081,Chipounov:2011:SPI:1950365.1950396}. There are many other use-cases, including security testing~\cite{Cha:2012:UMB:2310656.2310692,DBLP:conf/ndss/StephensGSDWCSK16,Pasareanu:2016:MSC} and complexity analysis~\cite{wise,Luckow2017}.

The research presented in this paper is based on \symbolicpathfinder (\spf), which extends the \javapathfinder framework with a symbolic execution mode \cite{Pasareanu:2013:ASE}. This mature symbolic execution tool works on \tool{Java ByteCode} and has support for most language features such as all primitive data types, strings, complex data structures, library calls, etc. It interfaces with a variety of solvers to solve constraints generated by symbolic execution.

\section{Approach}
\label{sec:approach}

\usetikzlibrary{patterns}
\tikzstyle{block} = [rectangle, draw, text width=4.5em, text centered, inner sep=2pt, minimum height=2em]
\tikzstyle{mediumblock} = [rectangle, draw, text width=7em, text centered, inner sep=2pt, minimum height=2em]
\tikzstyle{wideblock} = [rectangle, draw, text width=9em, text centered, inner sep=2pt, minimum height=2em]
\tikzstyle{blockinsidehatchedblock} = [rectangle, draw=white, text width=2.5em, text centered, inner sep=2pt, minimum height=1em,fill=white]
\tikzstyle{hatchedblock} = [pattern=north west lines, pattern color=gray, rectangle, draw, text width=4.5em, text centered, inner sep=2pt, minimum height=3em]
\tikzstyle{smallblock} = [rectangle, draw, text width=4em, text centered, inner sep=2pt, minimum height=3em]
\tikzstyle{widedashedblock} = [rectangle, draw,dashed, text width=9em, text centered, inner sep=2pt, minimum height=2em]
\tikzstyle{arrow} = [draw, -latex]
\begin{figure}
{\footnotesize
\begin{tikzpicture}[scale=1, every node/.style={transform shape}]
\node [hatchedblock] at (0,0) (fuzzerblock) {};
\node [blockinsidehatchedblock]  at (0,0) (fuzzertext) {fuzzer};

\node [mediumblock] at (+1,-0.8) (import) {import inputs};
\node[circle,draw] at (+2.2,-0.8) {\tiny 1};
\node [mediumblock] at (-1,-0.8) (export) {export inputs};
\node[circle,draw] at (-2.2,-0.8) {\tiny 5};

\draw [color=gray](-1.5,-3.65) rectangle (1.5, -1.9);
\node at (1.5,-1.95) [below=0mm, right=0mm, left=0mm, anchor=west] {\textsc{trie extension /}};
\node at (1.5,-2.20) [below=0mm, right=0mm, left=0mm, anchor=west] {\textsc{input assessment}};
\node [wideblock] at (0,-2.30) (concolicexecution) {concolic execution};
\node [widedashedblock] at (0,-3.25) (wcaanalysis) {worst-case analysis};
\node at (0.05,-2.75) [below=0mm, right=0mm, left=0mm, anchor=west] {\tiny includes};
\node[circle,draw] at (+1.75,-2.55) {\tiny 2};

\draw [color=gray](0.5,-6.0) rectangle (3.5, -4.2);
\node at (2.1,-4.10) [below=0mm, right=0mm, left=0mm, anchor=west] {\textsc{exploration}};
\node [wideblock] at (2, -4.65) (trieguidedsymex) {trie-guided symbolic \\execution};
\node [wideblock] at (2, -5.55) (boundedsymex) {bounded symbolic \\execution};
\node[circle,draw] at (+3.75,-4.45) {\tiny 3};

\draw [color=gray](-3.5,-6.0) rectangle (-0.5, -4.2);
\node at (-2.6,-4.10) [below=0mm, right=0mm, left=0mm, anchor=west] {\textsc{input generation}};
\node [wideblock] at (-2,-4.65) (modelgeneration) {model generation};
\node [wideblock] at (-2,-5.55) (inputgenerator) {input generator};
\node[circle,draw] at (-3.75,-4.45) {\tiny 4};

\draw [color=gray,dotted,thick](-4.1,-6.14) rectangle (4.1, -1.7);
\node at (-4.1,-1.55) [below=0mm, right=0mm, left=0mm, anchor=west] {\textit{SymExe}};

\path [arrow] (0.63,0) -- +(0.4,0) -- (1.03,-0.56);
\path [arrow]  (-1.03,-0.56) -- +(0,+0.57) -- (-0.63,0);

\path [arrow] (1.03, -1.05) -- (1.03, -1.9); 
\node at (1.03,-1.3) [below=0mm, right=0mm, left=0mm, anchor=west] {\tiny Mode.IMPORT};

\path [arrow, dashed] (concolicexecution) -- (wcaanalysis);

\path [arrow]  (1,-3.65) -- (1,-4.2);
\node at (1.0,-3.85) [below=0mm, right=0mm, left=0mm, anchor=west] {\tiny \textit{most promising node}};

\path [arrow] (trieguidedsymex) -- (boundedsymex);

\path [arrow]  (0.5,-5.1) -- (-0.5,-5.1);

\path [arrow] (modelgeneration) -- (inputgenerator);

\path [arrow]  (-3.0,-4.2) -- +(0,1.4) -- (-1.5,-2.8);
\node at (-2.8,-3.0) [below=0mm, right=0mm, left=0mm, anchor=west] {\tiny Mode.EXPORT};

\path [arrow] (-1.03, -1.9) -- (-1.03, -1.05); 
\node at (-0.98,-1.3) [below=0mm, right=0mm, left=0mm, anchor=west] {\tiny \textit{interesting}};
\node at (-0.98,-1.5) [below=0mm, right=0mm, left=0mm, anchor=west] {\tiny \textit{input}};

\end{tikzpicture}
}
\vspace{-7pt}
\caption{\badger workflow. Dashed lines represent activities that happen in parallel to the main flow.}
\label{fig:badger-overview}
\end{figure}
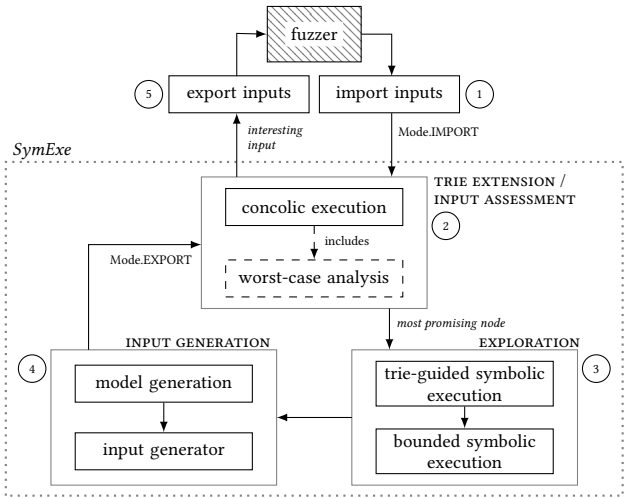

Figure \ref{fig:badger-overview} shows an overview of the \badger workflow.
The fuzzer block is hatched because the detailed workflow in the fuzzer is omitted in this figure. As discussed, \badger has two main components: fuzzing and symbolic execution (\symexe for short).
Inputs are generated in an iterative manner, on both fuzzing and \symexe sides, and are transferred between the two components, to trigger generation of new inputs, as guided by heuristics for worst-case analysis. Specifically,
the fuzzer generates and exports inputs that are shown to increase either coverage or cost on the fuzzer side; the specific costs will be discussed in detail below.
These inputs are imported by \symexe (cf. label 1 in Figure \ref{fig:badger-overview}).
\symexe uses concolic execution over the imported inputs to build a partial symbolic execution tree, similar to the trie-based data structure by Yang et al. \cite{Yang2012} (cf. label 2 in Figure \ref{fig:badger-overview}).
This trie captures the so-far explored portion of the symbolic execution paths, in which each node represents a decision in the program that includes symbolic variables.
The trie is saved in memory and gets extended incrementally, whenever new inputs are imported by \symexe.
While the trie is constructed and updated, it is also analyzed on-the-fly to compute the cost and coverage achieved for each node. This information is used by heuristics for worst-case analysis on the \symexe side to choose a node for further exploration in order to produce new inputs that increase coverage and/or cost. 

In order to explore the chosen node, we first use a guided symbolic execution until we reach the node of interest (cf. label 3 in Figure \ref{fig:badger-overview}).
This can be done very efficiently, guided by the path in the trie that leads to that node, and with constraint solving turned off.
After reaching the node of interest, we start a bounded symbolic execution (BSE) with a specified depth to explore new paths and generate corresponding new path conditions.

The collected path conditions are then solved to generate new input files (cf. label 4 in Figure \ref{fig:badger-overview}).
Since the exploration is performed on heuristically chosen nodes, and the newly generated inputs follow new paths, which were not explored before, 
we need to assess them, to measure their actual cost and the coverage achieved. This is done by running again concolic execution over these newly generated inputs, and updating/extending the trie in the process (cf. label 2 in Figure \ref{fig:badger-overview}).
Only the inputs that are found to lead to new interesting behavior (better cost or new coverage) are exported to the fuzzer, which will use them for its own analysis (cf. label 5 in Figure \ref{fig:badger-overview}). The fuzzer will generate more inputs of interest, which will be imported again by \symexe.

We group the steps labeled with 2 to 4 in the \symexe box, which represents the symbolic execution component, and does not include the interaction with the fuzzer.
We note that in practice, we let \badger stay in \symexe for a specified number of iterations, i.e. it only imports new input files from the fuzzer in intervals. The intuition is to let \symexe work on its own, exploring {\em several} promising nodes, rather than spending all its time importing information from the fuzzer. Note that similarly, the fuzzer imports new files in intervals (i.e. the havoc cycle in \afl).

\badger uses  \kelinci for fuzzing and \spf for symbolic execution, which are executed in parallel.
Both tools have been extended to search specifically for worst-case behavior w.r.t. a variety of cost metrics.
In the following, we explain both components in detail.


\subsection{Fuzzing with KelinciWCA}
\kelinciwca extends \kelinci with prioritization of costly paths. Costs are collected on the \java side, then sent to \afl, which we also modified to take into account the path costs (in addition to coverage). The fuzzer maintains the current \emph{highscore} with respect to the used cost model. 
When creating the next generation of inputs, the fuzzer selects ancestors from the set of inputs from the previous generation that either lead to the execution of previously unexplored program branches or to a new highscore. The chance that an input is selected from this set depends on its cost, as recorded on the \java side.

There are three cost models available:
\begin{itemize}
  \item \textbf{Timing} is measured by counting jumps (branches) in the program. This is more precise than measuring wall clock time, as in the latter case, there are often outliers and other inconsistencies due to, e.g., other activities on the machine. It is efficiently measured by adding the statement \prog{Mem.jumps++} to the instrumentation trampoline, adding 4 bytecode instructions and bringing the total number to 16.
  \item \textbf{Memory} usage is measured by intermittent polling using a timer. The maximum consumption at any point during execution of the program on the given input is collected. Though measuring allocations using program instrumentation could in some cases be more precise, it does not take into account garbage collection, and it requires the program to determine the sizes of individual objects which is expensive and can also be inaccurate.
  \item \textbf{User-defined} costs can also be used. In this case, the user instruments their program with calls to the special method \method{Kelinci.addCost(int)}, enabling the use of arbitrary metrics like the values of variables. Moreover, it allows a relationship between input and cost about which a machine can reason. This will be used later by the symbolic execution engine to directly generate inputs with maximal costs as an optimization.
\end{itemize}

\kelinciwca inherits the ability of \afl to run in a parallel mode, which enables the synchronization with other \afl instances.
After a configurable number of cycles with its own mutation operations (i.e. havoc cycles), \afl checks the other instances for interesting inputs.
Since this synchronization procedure is merely based on a specific folder structure, we can pass files from our symbolic execution part to \kelinciwca easily.

\subsection{Example}
Before describing the \symexe component of \badger, we introduce an example to illustrate how the various steps work.
The example is an implementation of Insertion Sort and is given in 
Listing~\ref{lst:insertionsort}.

\begin{lstlisting}[caption=Insertion Sort, label=lst:insertionsort, numbers=left,  stepnumber=1, xleftmargin=1em]
public static void sort(int[] a) {
    final int N = a.length;
    for (int i = 1; i < N; i++) {
        int j = i - 1;
        int x = a[i];
        while ((j >= 0) && (a[j] > x)) {
            a[j + 1] = a[j];
            j--;
        }
        a[j + 1] = x;
    }
}
\end{lstlisting}

\subsection{SymExe: Symbolic Execution with Symbolic PathFinder}
The \symexe component consists of three steps: trie extension and input assessment (cf. label 2 in Figure \ref{fig:badger-overview}), exploration (cf. label 3 in Figure \ref{fig:badger-overview}) and input generation (cf. label 4 in Figure \ref{fig:badger-overview}).
The following sub sections will cover all parts.

Figure~\ref{fig:example} shows snapshots of the trie while running \symexe on Insertion Sort for three numbers (N=3).
Note that trie nodes correspond to decisions rather than conditions in the program; multiple trie nodes may correspond to a single program condition. Since $N$ has the concrete value 3, the only symbolic decision in our program is \prog{a[j] > x} on line~5. Therefore, all nodes in Figure~\ref{fig:example} except the root nodes correspond to that decision.

\usetikzlibrary{patterns,shapes}
\tikzstyle{arrow} = [draw, -latex]
\tikzstyle{trienode} = [ellipse, draw, text width=4.5em, text centered, inner sep=2pt, minimum height=3em]
\tikzstyle{trienodeleaf} = [ellipse, draw, text width=4.5em, text centered, inner sep=2pt, minimum height=3em,fill=lightgray]
\tikzstyle{trienodehatched} = [pattern=north west lines, pattern color=gray, ellipse, draw, text width=4.5em, text centered, inner sep=2pt, minimum height=6.8em]
\tikzstyle{blockinsidehatchedblock} = [rectangle, draw=none, text width=4em, text centered, inner sep=2pt, minimum height=1em,fill=white]

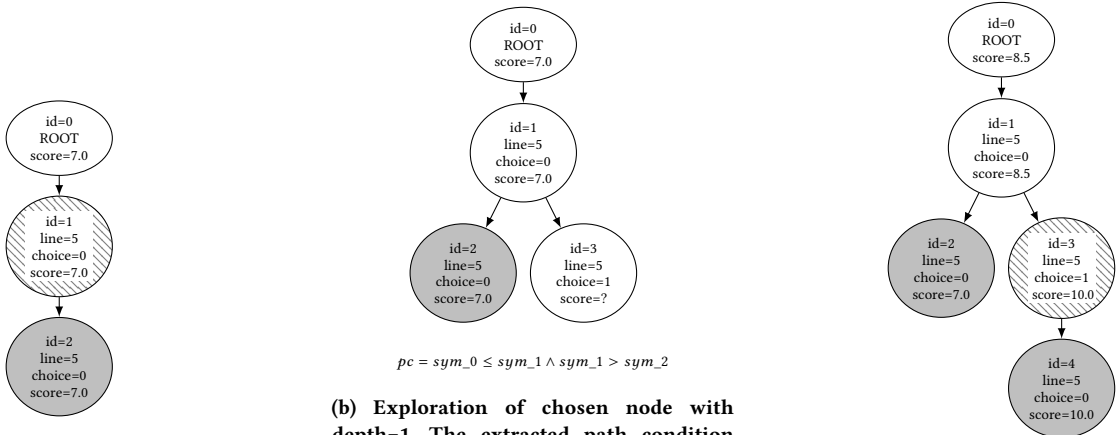
\begin{figure*}[t]
\centering
\begin{subfigure}[b]{0.3\textwidth}
\begin{center}
{\footnotesize
\begin{tikzpicture}[scale=\figstepstikzscale, every node/.style={transform shape}]
\node [trienode] at (0,-0.2) (root) {id=0 \\ROOT \\score=7.0};
\node [trienodehatched] at (0,-2) (n1) {};
\node [blockinsidehatchedblock]  at (0,-2) (n1text) {id=1 \\line=5 \\choice=0 \\score=7.0};
\node [trienodeleaf] at (0,-4) (n2) {id=2\\line=5 \\choice=0 \\score=7.0};
\path [arrow] (root) -- (n1);
\path [arrow] (n1) -- (n2);
\end{tikzpicture}
}
\vspace{.5cm}
\end{center}
\caption{Trie extension with initial input. The most promising node gets selected.}
\label{fig:example:a}
\end{subfigure}
\hfill
\begin{subfigure}[b]{0.3\textwidth}
\begin{center}
{\footnotesize
\begin{tikzpicture}[scale=\figstepstikzscale, every node/.style={transform shape}]
\node [trienode] at (0,-0.2) (root) {id=0 \\ROOT \\score=7.0};
\node [trienode] at (0,-2) (n1) {id=1 \\line=5 \\choice=0 \\score=7.0};
\node [trienodeleaf] at (-1,-4) (n2) {id=2\\line=5 \\choice=0 \\score=7.0};
\path [arrow] (root) -- (n1);
\path [arrow] (n1) -- (n2);
\node [trienode] at (1,-4) (n3) {id=3 \\line=5 \\choice=1 \\score=?};
\path [arrow] (n1) -- (n3);
\node at (2.5,-5.5) [below=0mm, right=0mm, left=0mm, text centered] {$pc = sym\_0 \leq sym\_1 \wedge sym\_1 > sym\_2$};  
\end{tikzpicture}
}
\end{center}
\caption{Exploration of chosen node with depth=1. The extracted path condition (pc) specifies input that reaches the node with id=3 (the input is an array \prog{[sym\_0, sym\_1, sym\_2]}).}
\label{fig:example:b}
\end{subfigure}
\hfill
\begin{subfigure}[b]{0.3\textwidth}
\begin{center}
{\footnotesize
\begin{tikzpicture}[scale=\figstepstikzscale, every node/.style={transform shape}]
\node [trienode] at (0,-0.2) (root) {id=0 \\ROOT \\score=8.5};
\node [trienode] at (0,-2) (n1) {id=1 \\line=5 \\choice=0 \\score=8.5};
\node [trienodeleaf] at (-1,-4) (n2) {id=2\\line=5 \\choice=0 \\score=7.0};
\path [arrow] (root) -- (n1);
\path [arrow] (n1) -- (n2);
\node [trienodehatched] at (1,-4) (n3) {};
\node [blockinsidehatchedblock]  at (1,-4) (n3text) {id=3 \\line=5 \\choice=1 \\score=10.0};
\path [arrow] (n1) -- (n3);
\node [trienodeleaf] at (1,-6) (n4) {id=4\\line=5 \\choice=0 \\score=10.0};
\path [arrow] (n3) -- (n4);
\end{tikzpicture}
}
\end{center}
\vspace{-4pt}
\caption{Assessment of generated input and extension of the trie. New most promising node gets selected.}
\label{fig:example:c}
\end{subfigure}
\caption{Trie evolution for Insertion Sort (N=3). The most promising node for the next exploration step is presented as hatched node. Grey colored nodes denote leafs, i.e., the last decision of an execution trace.}
\label{fig:example}
\end{figure*}

\subsubsection{Trie extension / input assessment}
In this step, \symexe performs a concolic execution over a set of concrete inputs (which are generated either by the fuzzer or by \symexe itself) and updates the trie data structure that maintains a view of all the explored symbolic execution paths. This is done by adding a concolic mode to SPF that simply collects the constraints along the concrete paths (without any constraint solving) and using a listener to monitor the execution and update the trie with newly explored nodes.

Figure \ref{fig:example:a} shows the trie for the initial input of the Insertion Sort example.
There are three nodes: the root node of the trie, and two decisions for \prog{a[j] > x} on line 5, which is the only condition that depends on symbolic input.
The last node (id=2) is a leaf, the last decision on the execution trace for the initial input.

During concolic execution we perform a worst-case analysis (WCA), to compute cost and coverage information for each node in the trie.
The default cost metric is timing, measured in number of jumps (branches).
Alternatively, WCA can collect user-defined costs 
specified in the code with \method{Observations.addCost(int)}. A novelty of our approach is that we can handle symbolic (input-dependent) costs as described in detail below.

Each trie node gets an associated \emph{score}, indicating how promising this node is with respect to the cost.
A score reflects an estimation of total costs that may be achieved rather than costs accumulated so far along a path. The score of a leaf node is defined as the cost associated with the path that ends in that leaf node; for example, it could be the total number of branches executed to reach that node.  The scores for the internal nodes
are propagated up in the trie, by taking the average of children scores as the score for a parent node.
%
%
%
For our simple example in Figure \ref{fig:example:a} all nodes get the value 7 (after importing one input) as the graph corresponds to a single path with cost 7.
Furthermore, coverage information for each node is updated.

The next step is to select the most promising node for further exploration. Every node that has unexplored children is a candidate for further exploration. We choose the most promising one based on three criteria. First, nodes with unexplored choices leading to new \textbf{branch} coverage are given priority; this means that if exploring a node can potentially lead to new coverage, we consider it as a good candidate for exploration. Second, nodes are ranked based on their \textbf{score}; nodes with higher score will again be given priority. Third, nodes are ranked based on their \textbf{position} in the trie. For the latter, the tool is parametric with respect to two different heuristics:
%
\begin{enumerate}
\item Prefer nodes \textbf{higher} in the trie. 
  This follows the intuition of exploring a wider region of the execution tree, covering a wider input space, analogous to a breadth-first search.
\item Prefer \textbf{lower} nodes. This follows the intuition to drill deeper in a promising direction, where high costs have been observed, analogous to depth-first search.
\end{enumerate}

For the example in Figure \ref{fig:example:a} we apply the second heuristic, i.e. to prefer nodes lower in the trie.
Hence, the node with id=1 is chosen as the most promising node 
instead of the root node.
For the situation in Figure \ref{fig:example:c} (which reflects the case when the import generated by \symexe itself is evaluated) it is clearly the node with id=3 because it is the only node with score=10.0 that can be explored.

We note that the input assessment and the trie extension can be executed in two modes: (1) import, and (2) export.
The import mode is used when importing input files from \kelinciwca. 
There is a special case when the user-defined cost metric is used: after importing an input, \symexe tries to maximize the costs along the concrete path by leveraging the maximizing procedures described in Section \ref{subsubsec:inputgeneration}.
If successful, \symexe will immediately export the maximizing input to the fuzzer.
The export mode is used when it is necessary to assess input files generated by \symexe.
Executing them with concolic execution (and on-the-fly WCA) reveals cost values and extends the trie with new nodes.
We use a conservative approach to exporting input files to \kelinciwca, because we do not want to flood the fuzzer with irrelevant inputs.
Only inputs leading to new branch coverage or a new high score are exported. 

\subsubsection{Exploration}
This step performs the actual exploration of new parts of the trie. The goal is to explore {\em new} branches starting from the node that was deemed most promising in the previous step (this means that the new branches will likely increase coverage or cost).
This step involves a quick symbolic execution along the trie path that reaches the node, as guided by the choices encoded in the trie.
There is no constraint solving needed.
As soon as the node of interest is reached, we switch to bounded symbolic execution (BSE) to explore new decisions.
When BSE reaches its bound we extract the path condition(s) and pass them to the next step, which will solve them to obtain new inputs.

In our example in Figure~\ref{fig:example:b} the trie-guided symbolic execution phase is very short, since only one choice is made to get to the node with id=1.
Next, we perform BSE with depth=1, i.e. only one step, and reach the node with id=3.
The score for this new node is unknown because it is not the end of the execution trace. The extracted path condition $pc$ makes it possible to generate an input value that reaches this new node. We will then need to run concolic execution again, to assess this new input, to compute the scores and update the trie (see Figure~\ref{fig:example:c}).

\subsubsection{Input Generation and Input-dependent Costs}
\label{subsubsec:inputgeneration}
In this step we generate concrete inputs by solving the path conditions from the previous step. This is done by using an off-the-shelve constraint solver. For the example, the path condition $sym\_0 \leq sym\_1  \wedge  sym\_1 > sym\_2$ may be solved to generate the input \prog{[1,1,0]}.
The solution(s) are used to generate a new input file (cf. step 5 in Figure~\ref{fig:badger-overview}), which is application-dependent. Note that it could happen that some path conditions are unsatisfiable, in which case no input is generated for them.

The path condition from Figure~\ref{fig:example:b} makes it possible to generate an input value that follows a path along the node with id=3. As shown in Figure~\ref{fig:example:c}, this input leads to a leaf node with id=4, and to a new high score, which is back-propagated to the precedent nodes.

As mentioned, the usage of heuristics makes it necessary to assess the actual cost value of each input because it is not guaranteed that the most promising nodes actually lead to worse costs. 
This is done by passing the generated input back to the trie extension phase and executing in export mode as described above.

A key novelty of our approach is the incorporation of symbolic (input-dependent) costs, which require specialized techniques for input generation. The user-specified costs allow to use arbitrary variables in the program when specifying the cost, including input-dependent ones. Thus, a symbolic path may have a symbolic cost, i.e. a symbolic expression that is passed as a parameter to method \method{Observations.addCost(int)}. We are then interested in computing inputs that {\em maximize} this cost. To achieve this, we propose to use an optimization procedure in the constraint solver.
In our set-up, we use \tool{Z3}~\cite{DeMoura:2008:ZES:1792734.1792766}, which allows to specify {\em optimization objectives}, i.e. we can ask \tool{Z3} to generate a model that maximizes a given expression. We illustrate the approach on a simple example, shown in Listing~\ref{lst:ex-maximization}.

\begin{lstlisting}[caption={User-Defined Cost Maximization Example}, label={lst:ex-maximization}, numbers=left,  stepnumber=1, xleftmargin=1em]
int sumArg(int[] a) {
	int sum = 0;
	for (int i=0; i < a.length; i++) {
		if (a[i] > 0) 
			sum += a[i];
		else 
 			sum -= a[i];
	}
	Observations.addCost(sum);
	return sum;
}
\end{lstlisting}

If the user is interested in maximizing the value of a variable (here \prog{sum}), then simply counting jumps or measuring resource consumption will not be sufficient for the generation of worst-case inputs.  
To address this situation, we instrument the code with a special cost specification (line~8).
When performing concolic execution over \prog{sumArg} for concrete input values \prog{a=\{1,2\}},  
variable \prog{sum} at line~8 has the value \prog{sum=$s_1$+$s_2$}  and the path condition is $s_1>0 \wedge s_2>0$, 
where $s_1$ and $s_2$ are the corresponding symbolic inputs.

We can pass an optimization term to \tool{Z3} for the specified cost expression.
For example for two positive inputs the query to the solver (in SMT 2 syntax) will look like:
{\small
\begin{center}
\prog{(assert (> $s_1$ 0))} \\
\prog{(assert (> $s_2$ 0))} \\
\prog{(maximize (+ $s_1$ $s_2$))}
\end{center}
}
The retrieved model for the path condition 
will contain values that maximize the given expression. Assume for simplicity that the allowed range for the inputs is $[-100,100]$.
Then the maximization procedure will return $s_1=100, s_2=100$ which indicate the worst-case inputs for this path. 


\section{Evaluation}
\label{sec:evaluation}
In this section we present an evaluation of our implementation for \badger. 
In order to enable the replication of our experiments and to make our tool publicly available, we have released \badger and all evaluation artifacts on GitHub: \url{https://github.com/isstac/badger}.
We evaluate our approach to answer these research questions:

\begin{itemize}
\item[\textbf{RQ1:}] Since \badger combines fuzzing and symbolic execution, is it better than each part on their own in terms of:
	\begin{enumerate}[label=(\alph*)]
		\item Quality of the worst-case, and 
		\item Speed ? 
	\end{enumerate}
\item[\textbf{RQ2:}] Is \kelinciwca better than \kelinci in terms of:
		\begin{enumerate}[label=(\alph*)]
		\item Quality of the worst-case, and 
		\item Speed ? 
	\end{enumerate}
\item[\textbf{RQ3:}] Can \badger reveal worst-case vulnerabilities?
\end{itemize}

\subsection{Experimental Setup}
\label{subsec:experimental-setup}

\paragraph{Subjects}
\label{par:subjects}

Table~\ref{tbl:subjects} gives an overview of our evaluation subjects.
Subject 1 to 5 are similar to the benchmarks used in \tool{SlowFuzz}. 
Subject 6 represents an image processing application provided by DARPA, as part of STAC engagements~\cite{Darpa}.
Subject 7 represents an implementation of a smart contract for crypto-currency, translated into \java from \tool{Ethereum}, where the goal is to estimate the worst-case {\em gas}
consumption.
For subject 1 to 6 we use the number of jumps as cost metric.
For subject 7 we use user-defined costs that are specified directly in the code.

\begin{table}[h]
\begin{center}
\caption{Overview of the evaluation subjects.}
\begin{tabular}{ l | l || l | l }
\textbf{ID} & \textbf{Subject} & \textbf{ID} & \textbf{Subject} \\ \hline
  1 & Insertion Sort & 4 & Hash Table \\
  2 & Quicksort & 5 & Compression \\ 
  3a & RegEx (fixed input) & 6 & Image Processor \\
  3b & RegEx (fixed regex) & 7 & Smart Contract \\
\end{tabular}
\label{tbl:subjects}
\end{center}
\end{table}

\paragraph{Experiment Execution}
\label{par:experiment-execution}

For all subjects we ran four variations: (1) \badger, (2) \kelinciwca, (3) \kelinci and (4) \symexe.
Success of an evolutionary fuzzing tool can depend greatly on the corpus of input files provided by the user as the first generation. For all experiments, we chose a single file with meaningless values (e.g. ``Hello World'') to leave the heavy lifting to the tools.
%
Running option (1) means to execute \kelinciwca and \symexe in parallel.
\kelinciwca starts with the initial input and \symexe imports the inputs from \kelinciwca like shown in Figure \ref{fig:badger-overview}.
Running option (2) and (3) means simply to execute the tools, and option (4) means to execute only \symexe on the initial input.
We have observed in pre-experiment executions that, for our subjects, after 5 hours the compared tools reach a plateau.
We therefore ran our experiments for 5 hours, 5 times (deviations from this paradigm for particular experiments are explained in the corresponding sections).
We also used these pre-experiment executions to determine the best fitting heuristic for each application.
Similar to Petsios et al.~\cite{Petsios2017}, we report slowdown in the subjects, i.e. the ratio between the costs of the observed worst case input and the initial input.

\paragraph{Infrastructure}
\label{par:infrastructure}
All experiments were conducted on a machine running \tool{openSUSE Leap 42.3} featuring 8 Quad-Core-AMD 8384 2.7 GHz and 64 GB of memory.
We used \openjdk 1.8.0\_151 and \tool{GCC 4.8.5}.
We configured the \tool{Java VM} to use at most 10 GB of memory.

\subsection{Sorting}
We first evaluate our approach on two textbook examples: Insertion Sort and Quicksort.
We use the implementations from \jdk 1.5. 
Results for Insertion Sort are the averages of 5 runs. For quicksort we deviated from the usual paradigm of 5 runs and performed 10 runs, because we observed a lot of variation in the first 5 runs.
For both subjects we used N=64, i.e. the input is an array of 64 integers.

\begin{figure}[h]
\begin{tikzpicture}[scale=\plotscale]
\begin{axis}[
  xlabel=time (minutes),
  ylabel= costs (\# jumps),
  xmajorgrids=true,  
  ymajorgrids=true,
  grid style=dashed,
  xmin=0, xmax=300,
  ymin=0, 
  x label style={at={(axis description cs:0.5,0.03)}},
  y label style={at={(axis description cs:0.06,0.5)}},
  width = \columnwidth,
  legend style={font=\footnotesize,at={(0.64,0.45)},anchor=west},
   ]
\addplot[color=red,mark=none, thick] table [y=badger, x=minutes]{01_insertionsort_updated.dat};
\addlegendentry{\badger}
\addplot[color=blue,mark=none,dotted, thick] table [y=kelinciwca, x=minutes]{01_insertionsort_updated.dat};
\addlegendentry{\kelinciwca}
\addplot[color=darkgreen,mark=none,dashed,thick] table [y=kelinci, x=minutes]{01_insertionsort_updated.dat};
\addlegendentry{\kelinci}
\addplot[color=brown,mark=none,dashdotted,thick] table [y=spf, x=minutes]{01_insertionsort_updated.dat};
\addlegendentry{\symexe}
\end{axis}
\end{tikzpicture}
\vspace{-10pt}
\caption{Results for Insertion Sort (N=64). }
\label{fig:insertionsort}
\end{figure}
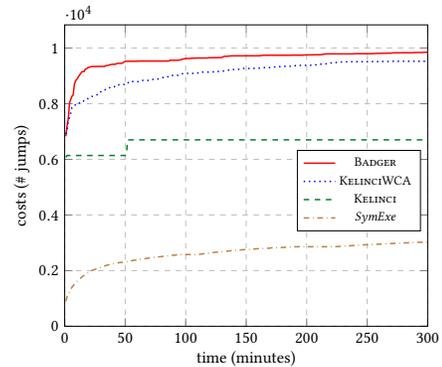
\vspace{2pt}

Results for Insertion Sort are shown in Figure~\ref{fig:insertionsort}. The score for the initial input is 509. 
The tools reach the following averaged final scores: \badger 9850, \kelinciwca 9533, \kelinci 6701, and \symexe 3025.
\kelinciwca produces a slowdown of 18.73x.
\badger reaches the final score of \kelinciwca already after 61 minutes (\kelinciwca needs 219 minutes) and continues to improve to a slowdown of 19.35x after 5 hours.
It is thus better in terms of both quality of the result, and speed.
The symbolic execution component \symexe by itself performs poorly, since it is only executed on the initial input and cannot use intermediate results by \kelinciwca.


\begin{figure}[h]
\begin{tikzpicture}[scale=\plotscale]
\begin{axis}[
  xlabel=time (minutes),
  ylabel= costs (\# jumps),
  x label style={at={(axis description cs:0.5,0.03)}},
  y label style={at={(axis description cs:0.02,0.5)}},
  xmajorgrids=true,  
  ymajorgrids=true,
  grid style=dashed,
  xmin=0, xmax=300,
  ymin=0, 
  width = \columnwidth,
  legend style={font=\footnotesize,at={(0.64,0.3)},anchor=west},
   ]
\addplot[color=red,mark=none, thick] table [y=badger, x=minutes]{02_quicksort_updated.dat};
\addlegendentry{\badger}
\addplot[color=blue,mark=none,dotted, thick] table [y=kelinciwca, x=minutes]{02_quicksort_updated.dat};
\addlegendentry{\kelinciwca}
\addplot[color=darkgreen,mark=none,dashed,thick] table [y=kelinci, x=minutes]{02_quicksort_updated.dat};
\addlegendentry{\kelinci}
\addplot[color=brown,mark=none,dashdotted,thick] table [y=spf, x=minutes]{02_quicksort_updated.dat};
\addlegendentry{\symexe}
\end{axis}
\end{tikzpicture}
\vspace{-10pt}
\caption{Results for Quicksort (N=64). }
\label{fig:quicksort}
\end{figure}
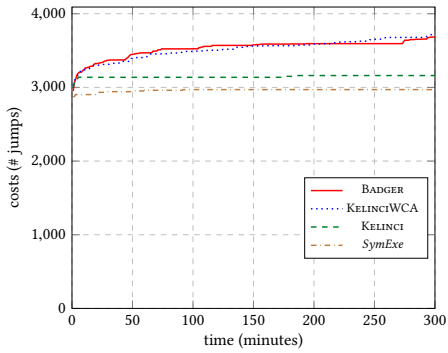

Results for Quicksort are shown in Figure~\ref{fig:quicksort}. The score for the initial input is 2829.
The tools reach the following averaged final scores: \badger 3683, \kelinciwca 3719, \kelinci 3161, and \symexe 2970.
There is no significant difference between the results for \badger and \kelinciwca.
\badger tends to be faster in generating a highscore between 20 and 150 minutes, but the final score after 300 minutes for \kelinciwca is slightly better.
\badger produces a slowdown of 1.30x and \kelinciwca of 1.31x.
This minor difference at the end can be explained by the randomness inherent in fuzzing; since \badger includes \kelinciwca, its results for a particular run are always at least as good. 
In fact for the best run of the 10 performed, we observed that \badger produced the score 4219 (1.49x), while \kelinciwca produced the score 4013 (1.42x).

\subsection{Regular Expressions}
The second evaluation considers regular expression matching, which can be vulnerable to so called ReDoS (Regular expression DoS) attacks.
Specifically, we used the \prog{java.util.regex} \jdk package.

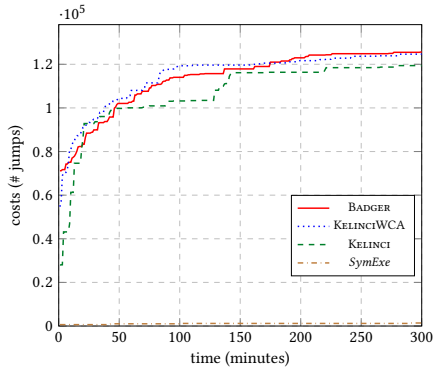
\begin{figure}[h!]
\begin{tikzpicture}[scale=\plotscale]
\begin{axis}[
  xlabel=time (minutes),
  ylabel= costs (\# jumps),
  xmajorgrids=true,  
  ymajorgrids=true,
  grid style=dashed,
  xmin=0, xmax=300,
  ymin=0, 
  x label style={at={(axis description cs:0.5,0.03)}},
  y label style={at={(axis description cs:0.06,0.5)}},
  width = \columnwidth,
  legend style={font=\footnotesize,at={(0.64,0.3)},anchor=west},
   ]
\addplot[color=red,mark=none, thick] table [y=badger, x=minutes]{03_regex_lorem_updated.dat};
\addlegendentry{\badger}
\addplot[color=blue,mark=none,dotted, thick] table [y=kelinciwca, x=minutes]{03_regex_lorem_updated.dat};
\addlegendentry{\kelinciwca}
\addplot[color=darkgreen,mark=none,dashed,thick] table [y=kelinci, x=minutes]{03_regex_lorem_updated.dat};
\addlegendentry{\kelinci}
\addplot[color=brown,mark=none,dashdotted,thick] table [y=spf, x=minutes]{03_regex_lorem_updated.dat};
\addlegendentry{\symexe}
\end{axis}
\end{tikzpicture}
\vspace{\belowplotvspace}
\caption{Results for Regular Expression (fixed regex). }
\label{fig:regex-text}
\end{figure}

We performed two experiments. In the first, we fixed the matching text and mutated the regular expression.
We used the lorem ipsum filler text, and limited mutated regular expressions to 10 characters.
As initial input we used the regular expression $\left[\symbol{92}s\symbol{92}S\right]^{*}$. 
We increased the number of experiments to 10 because we observed too much variation in the first 5 runs.

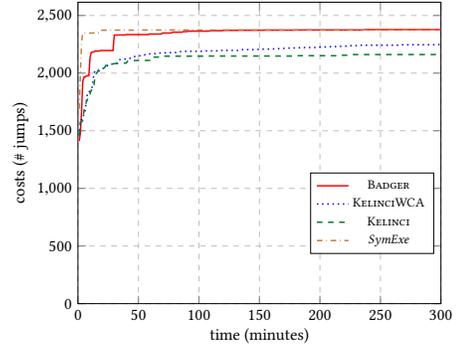
\begin{figure}[h]
\begin{tikzpicture}[scale=\plotscale]
\begin{axis}[
  xlabel=time (minutes),
  ylabel= costs (\# jumps),
  xmajorgrids=true,  
  ymajorgrids=true,
  grid style=dashed,
  xmin=0, xmax=300,
  ymin=0,
  x label style={at={(axis description cs:0.5,0.03)}},
  y label style={at={(axis description cs:0.02,0.5)}},
  width = \columnwidth,
  legend style={font=\footnotesize,at={(0.64,0.3)},anchor=west},
   ]
\addplot[color=red,mark=none, thick] table [y=badger, x=minutes]{03_regex_01_username_updated.dat};
\addlegendentry{\badger}
\addplot[color=blue,mark=none,dotted, thick] table [y=kelinciwca, x=minutes]{03_regex_01_username_updated.dat};
\addlegendentry{\kelinciwca}
\addplot[color=darkgreen,mark=none,dashed,thick] table [y=kelinci, x=minutes]{03_regex_01_username_updated.dat};
\addlegendentry{\kelinci}
\addplot[color=brown,mark=none,dashdotted,thick] table [y=spf, x=minutes]{03_regex_01_username_updated.dat};
\addlegendentry{\symexe}
\end{axis}
\end{tikzpicture}
\vspace{\belowplotvspace}
\caption{Results for Regular Expression (username).}
\label{fig:regex-username}
\end{figure}
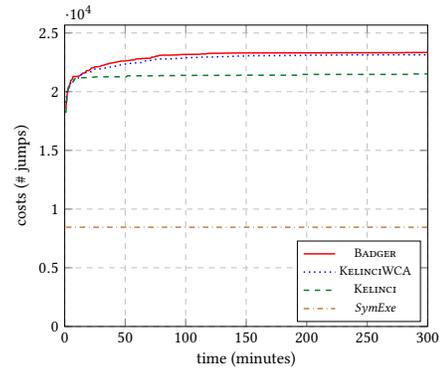
\begin{figure}[h]
\begin{tikzpicture}[scale=\plotscale]
\begin{axis}[
  xlabel=time (minutes),
  ylabel= costs (\# jumps),
  xmajorgrids=true,  
  ymajorgrids=true,
  grid style=dashed,
  xmin=0, xmax=300,
    ymin=0,
  x label style={at={(axis description cs:0.5,0.03)}},
  y label style={at={(axis description cs:0.06,0.5)}},
  width = \columnwidth,
  legend style={font=\footnotesize,at={(0.64,0.15)},anchor=west},
   ]
\addplot[color=red,mark=none, thick] table [y=badger, x=minutes]{03_regex_02_password_updated.dat};
\addlegendentry{\badger}
\addplot[color=blue,mark=none,dotted, thick] table [y=kelinciwca, x=minutes]{03_regex_02_password_updated.dat};
\addlegendentry{\kelinciwca}
\addplot[color=darkgreen,mark=none,dashed,thick] table [y=kelinci, x=minutes]{03_regex_02_password_updated.dat};
\addlegendentry{\kelinci}
\addplot[color=brown,mark=none,dashdotted,thick] table [y=spf, x=minutes]{03_regex_02_password_updated.dat};
\addlegendentry{\symexe}
\end{axis}
\end{tikzpicture}
\vspace{\belowplotvspace}
\caption{Results for Regular Expression (password).}
\label{fig:regex-password}
\end{figure}
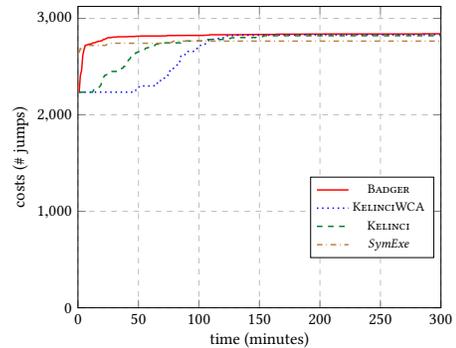
\begin{figure}[h]
\begin{tikzpicture}[scale=\plotscale]
\begin{axis}[
  xlabel=time (minutes),
  ylabel= costs (\# jumps),
  xmajorgrids=true,  
  ymajorgrids=true,
  grid style=dashed,
  xmin=0, xmax=300,
  ymin=0,
    x label style={at={(axis description cs:0.5,0.03)}},
  y label style={at={(axis description cs:0.02,0.5)}},
  width =\columnwidth,
  legend style={font=\footnotesize,at={(0.64,0.3)},anchor=west},
   ]
\addplot[color=red,mark=none, thick] table [y=badger, x=minutes]{03_regex_03_hexcolor_updated.dat};
\addlegendentry{\badger}
\addplot[color=blue,mark=none,dotted, thick] table [y=kelinciwca, x=minutes]{03_regex_03_hexcolor_updated.dat};
\addlegendentry{\kelinciwca}
\addplot[color=darkgreen,mark=none,dashed,thick] table [y=kelinci, x=minutes]{03_regex_03_hexcolor_updated.dat};
\addlegendentry{\kelinci}
\addplot[color=brown,mark=none,dashdotted,thick] table [y=spf, x=minutes]{03_regex_03_hexcolor_updated.dat};
\addlegendentry{\symexe}
\end{axis}
\end{tikzpicture}
\vspace{\belowplotvspace}
\caption{Results for Regular Expression (hexcolor).}
\label{fig:regex-hexcolor}
\end{figure}

Results are shown in Figure~\ref{fig:regex-text}. The initial input score is 68101.
The tools reach the following average scores: \badger 125616, \kelinciwca 124641, \kelinci 119393, and \symexe 1388.
\badger produces a slowdown of 1.84x and \kelinciwca of 1.83x.
This insignificant difference can be explained by the poor result for \symexe, which was not able to improve the score of the initial input.

In the second experiment, we fix the regex and mutate the text.
We use the lorem ipsum as initial input again.
For the regular expressions we use ten popular examples~\cite{regular_expressions}.
Due to space limitations, we include only the first three here, respectively matching a username, password, and hexadecimal color code:

\vspace{-5pt}
{\Small
\begin{equation}
\hat{}\left[a\!-\!z0\!-\!9\_\!-\!\right]\{3,15\}\$
\end{equation}
\begin{equation}
((?\!=\!.\!^{*}\!\symbol{92}d)(?\!=\!.\!^{*}\!\left[a-z\right])(?\!=\!.\!^{*}\!\left[A-Z\right])(?\!=\!.\!^{*}\!\left[@\#\$\%\right]).\{6,20\})
\end{equation}
\begin{equation}
\hat{}\,\#(\left[A\!-\!Fa\!-\!f0\!-\!9\right]\!\{6\}\,|\,\left[A\!-\!Fa\!-\!f0\!-\!9\right]\!\{3\})\$
\end{equation}
}

Results are shown in Figures~\ref{fig:regex-username}, \ref{fig:regex-password} and \ref{fig:regex-hexcolor}, respectively.
Remarkably, for the username regex, \symexe is faster than \badger. 
This can be explained by the fact that the initial input already leads to relatively high costs; in general this is not the case.
Additionally, \symexe starts working right away, while \badger needs some time to import input files generated by \kelinciwca (it is started with a slight delay).

Results for the password regex show that \symexe is not able to generate any comparable highscore, which explains why \badger is not performing significantly better than \kelinciwca.

For the color code regex, \badger and \kelinciwca both produce a slowdown of 1.27x, but \badger finds it significantly faster.
By leveraging the inputs generated by \symexe, \badger reaches a cost of 2800 after 25 minutes, for which \kelinciwca needs 133 minutes.
Interestingly, \kelinci is also faster than \kelinciwca. 
While statistically unlikely, this can happen due to the inherent randomness in the fuzzer.

\subsection{Hash Table}
The fourth evaluation subject is a hash table implementation taken from a recent \darpa engagement \cite{Darpa} and modified to match the hash function by \tool{SlowFuzz}\cite{Petsios2017}, which was taken from a vulnerable \tool{PHP} implementation~\cite{phpvuln}.
The size of the hash table is 64, each key in the hash table has a length of 8 characters, and we fill it by reading the first $64 \cdot 8$ characters from an input file.
The worst-case of a hash table implementation can be triggered by generating hash collisions. Therefore, besides the normal costs, we also report the number of hash collisions. 
We executed the experiments 10 times because we observed too much variation in the first 5 runs.

\begin{figure}[h]
\begin{tikzpicture}[scale=\plotscale]
\begin{axis}[
  xlabel=time (minutes),
  ylabel= costs (\# jumps),
  xmajorgrids=true,  
  ymajorgrids=true,
  grid style=dashed,
  xmin=0, xmax=300,
  ymin=0,
    x label style={at={(axis description cs:0.5,0.03)}},
  y label style={at={(axis description cs:0.02,0.5)}},
  width = \columnwidth,
  legend style={font=\footnotesize,at={(0.64,0.2)},anchor=west},
   ]
\addplot[color=red,mark=none, thick] table [y=badger, x=minutes]{04_hashtable_updated.dat};
\addlegendentry{\badger}
\addplot[color=blue,mark=none,dotted, thick] table [y=kelinciwca, x=minutes]{04_hashtable_updated.dat};
\addlegendentry{\kelinciwca}
\addplot[color=darkgreen,mark=none,dashed,thick] table [y=kelinci, x=minutes]{04_hashtable_updated.dat};
\addlegendentry{\kelinci}
\addplot[color=brown,mark=none,dashdotted,thick] table [y=spf, x=minutes]{04_hashtable_updated.dat};
\addlegendentry{\symexe}
\end{axis}
\end{tikzpicture}
\vspace{\belowplotvspace}
\caption{Results for Hash Table (N=64, key length=8).}
\label{fig:hashtable}
\end{figure}
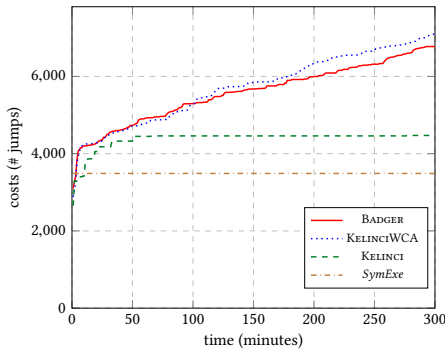

The graph in Figure \ref{fig:hashtable} shows that \badger first performs slightly better and faster, but is passed by \kelinciwca after 103 minutes.
This is based on \symexe, which quickly generates a very good score, but is not able to further improve.
Since we cannot report the number of collisions for the averaged plot, we looked at the best performing run for each experiment. 
The results correspond to 31 collisions found by \badger, and 39 found by \kelinciwca (the theoretical upper bound is 63).
This subject shows very evidently the advantage of \kelinciwca over \kelinci, which plateaus after 66 minutes and only finds 22 collisions.

\subsection{Compression}
Our fifth evaluation subject is taken from \commonscompress.
In the experiment, we \tool{BZip2} compress files up to 250 bytes. 

\begin{figure}[th]
\begin{tikzpicture}[scale=\plotscale]
\begin{axis}[
  xlabel=time (minutes),
  ylabel= costs (\# jumps),
  xmajorgrids=true,  
  ymajorgrids=true,
  grid style=dashed,
  xmin=0, xmax=300,
  ymin=0,
    x label style={at={(axis description cs:0.5,0.03)}},
  y label style={at={(axis description cs:0.06,0.5)}},
  width = \columnwidth,
  legend style={font=\footnotesize,at={(0.64,0.3)},anchor=west},
   ]
\addplot[color=red,mark=none, thick] table [y=badger, x=minutes]{05_compress_updated.dat};
\addlegendentry{\badger}
\addplot[color=blue,mark=none,dotted, thick] table [y=kelinciwca, x=minutes]{05_compress_updated.dat};
\addlegendentry{\kelinciwca}
\addplot[color=darkgreen,mark=none,dashed,thick] table [y=kelinci, x=minutes]{05_compress_updated.dat};
\addlegendentry{\kelinci}
\addplot[color=brown,mark=none,dashdotted,thick] table [y=spf, x=minutes]{05_compress_updated.dat};
\addlegendentry{\symexe}
\end{axis}
\end{tikzpicture}
\vspace{\belowplotvspace}
\caption{Results for Compression (N=250).}
\label{fig:compress}
\end{figure}
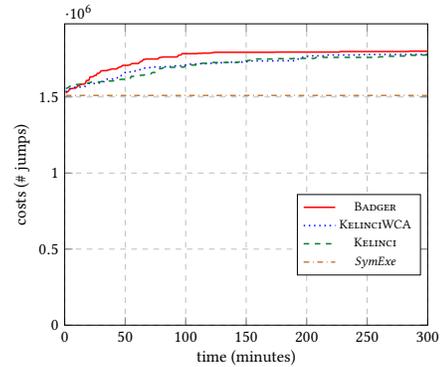

Results are shown in Figure~\ref{fig:compress}. 
The score for the initial input is 1,505,039.
The tools reach the following average scores: \badger 1,800,831, \kelinciwca 1,779,457, \kelinci 1,775,438, and \symexe 1,509,880.
\badger produces a slowdown of 1.20x and \kelinciwca of 1.18x.
\badger is significantly faster.
The worst case found by \kelinciwca after 100 minutes is found by \badger within 50 minutes.

\subsection{Image Processor}
\label{subsec:imageprocessor}

Our sixth evaluation subject is an image processing application taken from a recent \darpa engagement \cite{Darpa}. 
Our analysis revealed a vulnerability related to particular pixel values in the input image causing a significantly increased runtime for the program.
These pixel values trigger a particular value from a static array, which is used to determine the number of iterations in a processing loop.
\badger was able to automatically generate a JPEG image that exposes the vulnerability.
For the sake of simplicity, we limited the size of images to 2x2 pixels.

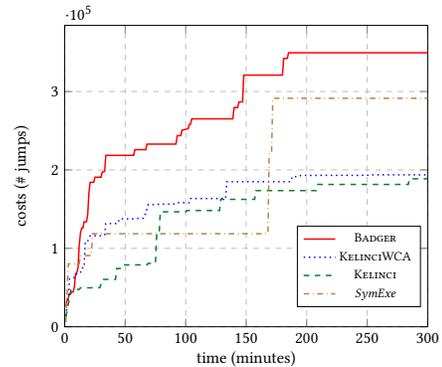
\begin{figure}[h]
\begin{tikzpicture}[scale=\plotscale]
\begin{axis}[
  xlabel=time (minutes),
  ylabel= costs (\# jumps),
  xmajorgrids=true,  
  ymajorgrids=true,
  grid style=dashed,
   xmin=0, xmax=300,
  ymin=0,
    x label style={at={(axis description cs:0.5,0.03)}},
  y label style={at={(axis description cs:0.06,0.5)}},  
  width = \columnwidth,
  legend style={font=\footnotesize,at={(0.64,0.2)},anchor=west},
   ]
\addplot[color=red,mark=none, thick] table [y=badger, x=minutes]{06_imageprocessor_updated.dat};
\addlegendentry{\badger}
\addplot[color=blue,mark=none,dotted, thick] table [y=kelinciwca, x=minutes]{06_imageprocessor_updated.dat};
\addlegendentry{\kelinciwca}
\addplot[color=darkgreen,mark=none,dashed,thick] table [y=kelinci, x=minutes]{06_imageprocessor_updated.dat};
\addlegendentry{\kelinci}
\addplot[color=brown,mark=none,dashdotted,thick] table [y=spf, x=minutes]{06_imageprocessor_updated.dat};
\addlegendentry{\symexe}
\end{axis}
\end{tikzpicture}
\vspace{\belowplotvspace}
\caption{Results for Image Processor (2x2 JPEG).}
\label{fig:imageprocessor}
\end{figure}

Results are shown in Figure~\ref{fig:imageprocessor}. Here we see that \badger clearly outperforms both its components. It produces a slowdown of 40.11x, corresponding to theoretical worst-case, where all pixels have a value triggering the highest possible number of iterations in the processing loop.

\subsection{Smart Contract}

Our last subject is an implementation of a smart contract for cryptocurrency, where the goal is to analyse the usage of a resource called {\em gas} of \tool{Ethereum} software.
Exceeding the allocated budget could result in loss of cryptocurrency.
Therefore we consider gas as the user-defined cost in our analysis.
We manually instrument the code with calls to the special methods \method{Kelinci.addCost(int)}, and \method{Observations.addCost(int)}.
We executed the experiments 10 times because we observed too much variation in the first 5 runs.
For our experiments we used an input array size of $N=50$ items.

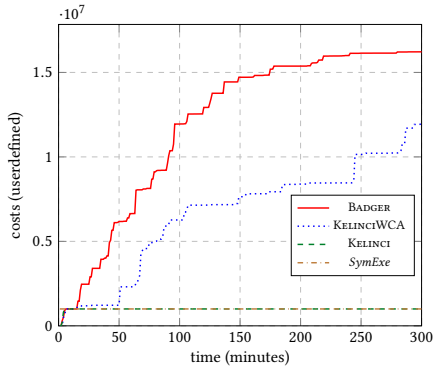
\begin{figure}[h]
\begin{tikzpicture}[scale=\plotscale]
\begin{axis}[
  xlabel=time (minutes),
  ylabel= costs (userdefined),
  xmajorgrids=true,  
  ymajorgrids=true,
  grid style=dashed,
   xmin=0, xmax=300,
  ymin=0,
    x label style={at={(axis description cs:0.5,0.03)}},
  y label style={at={(axis description cs:0.06,0.5)}},  
  width = \columnwidth,
  legend style={font=\footnotesize,at={(0.64,0.3)},anchor=west},
   ]
\addplot[color=red,mark=none, thick] table [y=badger, x=minutes]{07_smartcontract_updated.dat};
\addlegendentry{\badger}
\addplot[color=blue,mark=none,dotted, thick] table [y=kelinciwca, x=minutes]{07_smartcontract_updated.dat};
\addlegendentry{\kelinciwca}
\addplot[color=darkgreen,mark=none,dashed,thick] table [y=kelinci, x=minutes]{07_smartcontract_updated.dat};
\addlegendentry{\kelinci}
\addplot[color=brown,mark=none,dashdotted,thick] table [y=spf, x=minutes]{07_smartcontract_updated.dat};
\addlegendentry{\symexe}
\end{axis}
\end{tikzpicture}
\vspace{\belowplotvspace}
\caption{Results for Smart Contract (N=50).}
\label{fig:smartcontract}
\vspace{1em}
\end{figure}

Results are shown in Figure \ref{fig:smartcontract}.
We observed a cost of 3 for the initial input.
The tools reach the following average scores: \badger 16,218,905, \kelinciwca 11,934,664, \kelinci 1,000,107, and \symexe 1,000,107.
The increase in the cost is 5,406,301.67x for \badger and 3,978,221.33x for \kelinciwca.
The cost depends on a concrete value in the input and can thus be very large (even for a short path the cost can be large if the input value is large).
The initial input does not contain these large values, and hence, the cost increase is so dramatic.
\badger is significantly faster than \kelinciwca, and also produces a much higher worst case cost.

\subsection{Discussion} 

\subsubsection*{RQ1.a}
Our evaluation shows that, in terms of quality of the worst case, \badger is always better than \symexe because eventually \badger will use the insights of its symbolic execution part.
But, as mentioned earlier, \badger does not consider inputs from \symexe until they are imported by \kelinciwca.
Additionally, the randomness from the fuzzer can cause \symexe to explore other paths that turn out to be less costly than those closer to the initial input.
In most of our subjects, \badger also produces a better worst case than \kelinciwca.
There are two cases in which \kelinciwca is slightly better than \badger: Quicksort and Hash Table. These differences are based on the randomness in the fuzzing component of our approach.
Therefore, we conclude question RQ1.a with the positive answer: yes.
Note that in practice, a slightly lower worst case is often more useful if it can be obtained significantly faster.

\paragraph{RQ1.b}
Our evaluation demonstrates that \badger is significantly faster than \kelinciwca and \symexe, attesting a clear positive answer to RQ1.b.
In most cases, \symexe by itself performs poorly compared to \badger.
Nevertheless, there is one case where \symexe is able to generate a high score very fast, and \badger is not able to follow immediately: Regular Expression (username).
We explain this by the random initial input, along which path \symexe finds a high score very quickly, whereas \badger experiences some delay in importing \kelinciwca results that also sidetrack the analysis.
Note that, similar to \kelinciwca, \badger can achieve better performance through parallelization by, e.g., running multiple fuzzing instances in parallel.

\paragraph{RQ2.a}
The evaluation shows that \kelinciwca is always better than \kelinci in terms of quality of the worst case.
Although there are cases, in which \kelinci performs not bad, the single focus on coverage limits \kelinci for worst case analysis.
This limitation of \kelinci was eliminated with \kelinciwca.

\paragraph{RQ2.b}
Regarding the speed comparison between \kelinciwca and \kelinci, our evaluation illustrates that in almost all subjects \kelinciwca does not only retrieves a better worst case, but it also achieves this in shorter time.
The only exception is the subject Regular Expression (hexcolor).
In general, \kelinciwca is faster, but in cases where for a particular application higher coverage implies a better worst case, then \kelinci might be more efficient.

\paragraph{RQ3}
We have shown that Badger performed well in our experiments, finding slowdowns on the subjects. Notably, \badger identified a vulnerability in Image Processor, a complex application that processes non-trivial inputs (i.e. JPEG images). 
\badger was able to reveal the actual worst-case by building a JPEG image, demonstrating
its ability in exposing vulnerabilities in complex applications.

\subsection{Threats to Validity}

\paragraph{Internal Validity.}
The main threat to internal validity is the correctness of collection and analysis of experimental results.
Therefore, we fully automated the process of collecting data, aggregating values and plotting graphs.
Another threat to internal validity is the selection of experiment parameters, such as the heuristic worst case analysis.
In order to verify this we conduced pre-experimental tests that showed the effectiveness of our selection.
Additionally, in our experiments with \badger, we used one process for \kelinciwca and one for \symexe. In the experiments with the individual tools, each ran on a single core. This could give the combination of the tools an advantage. In future work, we plan to parallelize \symexe, which is expected to further improve the effectiveness of \badger, and also will enable a more fair comparison.

\paragraph{External Validity.}
The main threat to external validity is that evaluation subjects may not generalize.
In order to mitigate this threat we have selected benchmarks that match existing work in the field, and added a real-world (complex) example. 

\paragraph{Construct Validity.}
The main threat to construct validity is the correctness of our actual implementation.
We based our implementation on \spf and \kelinci, and hence, our adaptions inherit potential incorrectness of these tools.
However, face-validity shows that our evaluation results match the expected outcome.

\section{Related Work}
\label{sec:relwork}

\tool{SlowFuzz}~\cite{Petsios2017} is a fuzzer based on \tool{LibFuzzer}~\cite{libfuzzer} that prioritizes inputs that lead to increased execution times. The tool is similar to our \kelinciwca component, although it addresses a different programing language. Our evaluation is partially based on the \tool{SlowFuzz} evaluation and not surprisingly results obtained with \kelinciwca are similar to those for \tool{SlowFuzz}. Furthermore, our experiments with \badger indicate that a combination of \tool{SlowFuzz} with symbolic execution would achieve similar benefits.

Several tools explore the combination of symbolic execution with fuzzing. All these tools aim to increase coverage while our goal is to generate inputs that excersise behaviors that increase resource consumption, leading to significant technical differences. 

\tool{EvoSuite}~\cite{evosuite} is a test-case generation tool for \java, based on evolutionary algorithms and dynamic symbolic execution. Unit tests are generated via random mutation, recombination and selection and are evaluated with respect to a given fitness function (typically a coverage metric). When a change in fitness is observed after mutation of a certain primitive value, the variable this value is assigned to is deemed important and dynamic symbolic execution is invoked with this variable as a symbol. 

\tool{SAGE} (Scalable Automated Guided Execution)~\cite{Godefroid:2012:SWF:2090147.2094081} extends dynamic symbolic execution 
with a \emph{generational search} that, instead of negating only the final condition of a complete symbolic execution, negates all conditions on the path. Solving the resulting path conditions results in a large number of new test inputs. 
\tool{SAGE} is used extensively at Microsoft where it has been successful at finding many security-related bugs. 

\tool{Mayhem}~\cite{Cha:2012:UMB:2310656.2310692} is a symbolic execution engine that aims to find security vulnerabilities in binaries. 
%
A \emph{Concrete Executor Client} (CEC) explores paths concretely and performs a dynamic taint analysis. When a basic block is reached that contains tainted instructions, it is passed to the \emph{Symbolic Executor Server} (SES) that is running in parallel. After symbolic execution, the SES instructs the CEC on a particular path to execute. 
\tool{Mayhem} was combined with the \tool{Murphy} fuzzer and won the 2016 DARPA Cyber Grand Challenge~\cite{darpa_cgc}.

\tool{Driller}~\cite{DBLP:conf/ndss/StephensGSDWCSK16} is another promising tool that combines the \afl fuzzer with the \tool{angr} symbolic execution engine and that has achieved similar results to Mayhem. 
It is very similar to our approach, in that it executes a fuzzer and symbolic execution engine in parallel, combining their strengths and overcoming their weaknesses. However, while \tool{Driller} is optimized for uncovering new branches, we focus on worst-case analysis.

Symbolic execution was used before for worst-case analysis~\cite{wise,Luckow2017}. \tool{WISE}~\cite{wise} analyzes programs for small input configurations using concolic execution and attempts to {\em learn} a path policy that likely leads to worst-case executions at any size. This policy is then applied to programs that have larger input configurations, to {\em guide} the symbolic execution of the program. \tool{SPF-WCA}~\cite{Luckow2017} uses \spf to perform a similar analysis with more sophisticated path policies, which take into account the {\em history} of executions and the calling context of the analyzed procedures. In addition, \tool{SPF-WCA} also uses function fitting to obtain estimates of the asymptotic complexity. Both WISE and SPF-WCA require to perform exhaustive symbolic execution for large enough input sizes to obtain good policies, which may not be feasible in practice. Furthermore, both techniques only consider one input parameter for size and usually require some manual fine tuning (e.g. for \tool{SPF-WCA} manually decide the size of the history). 
In contrast, our technique uses a combination with fuzzing to avoid a full exhaustive symbolic execution and is fully automatic (except for the creation of drivers that are necessary for all of these approaches). Furthermore, \badger enables analysis with input-dependent costs, using a novel maximization technique. This significantly broadens the application scope of \badger w.r.t. previous techniques, which only support simple, fixed costs associated with each program path.

Load testing~\cite{LoadTesting} employs symbolic execution to perform an {\em iterative} analysis for increasing exploration depth, 
with pruning of paths with low resource consumption.
That work could not be used directly for finding the worst-case algorithmic behavior, 
since all the paths are explored up to the same depth, and therefore have the same number of steps. 

Probabilistic symbolic execution is used in~\cite{ICSE16} 
to infer the performance distribution of a program according to given usage profiles.
Although promising, the technique does not yet scale for large programs, due to
the more involved probabilistic computation.    

Static analysis is used in~\cite{Gulwani09,GulavaniGulwani08,AlbertAGP11a,ShkaravskaKerstenVanEekelen2010}
to compute conservative bounds on looping programs. In contrast to this work, our tool produces actual inputs that expose vulnerabilities, but can not provide guarantees on worst-case bounds. There is also a large body of work on worst-case execution time (WCET) analysis---in particular for real-time systems~\cite{malikbook,reinhardbook,automaticloopboundwcet,automaticloopbound2}. Unlike our work, these techniques typically assume that loops have finite bounds and are independent of input, and estimate worst-case execution for specific platforms. 

{\it Profilers}~\cite{gprof,SevitskyPK01, AmmonsCGS04} are typically used for performance analysis of programs, but are inherently limited by the quality of tests used. In our work we aim to automatically generate input data triggering a diverse set of executions, including the worst-case ones.

\section{Conclusions and Future Work}
\label{sec:conclusions}
We have proposed \badger, a hybrid testing approach for complexity analysis.
It extends the \kelinci fuzzer with a worst-case analysis, and uses a modified version of \symbolicpathfinder to import inputs from the fuzzer, analyze them and generate new inputs that increase both coverage and execution cost. \badger can use various cost models, such as time and memory consumption, and also supports the specification of input-dependent costs.
\badger was evaluated against a large set of benchmarks, demonstrating the performance and quality benefits over fuzzing and symbolic execution by themselves.

In the future, we plan to explore more heuristics for worst-case analysis on both the fuzzing and the symbolic execution side. We also plan to focus more on the symbolic execution part of \badger and conduct experiments using multiple depths during bounded symbolic execution.
Additionally, we plan to explore techniques to increase scalability of the symbolic execution part by, e.g., limiting the size of the trie, which would lead to a faster exploration of the upper symbolic execution tree.
Furthermore, we plan to extend our approach not only for complexity analysis, but also for a {\em differential} side-channel analysis of security-relevant applications. 
\begin{acks}
This material is based on research sponsored by DARPA under agreement number FA8750-15-2-0087.
The U.S. Government is authorized to reproduce and distribute reprints for Governmental purposes notwithstanding any copyright notation thereon.
This work is also supported by the German Research Foundation (GR 3634/4-1 EMPRESS).
\end{acks}

\newpage
\balance
\bibliographystyle{ACM-Reference-Format}
\bibliography{refs,bib-symex,refs1}

\end{document}